\documentclass[%
aip,
jcp,
amsmath,amssymb,
preprint,%
]{revtex4-2}

\usepackage{graphicx}
\usepackage{dcolumn}
\usepackage{bm}

\usepackage[utf8]{inputenc}
\usepackage[T1]{fontenc}
\usepackage{mathptmx}
\usepackage{etoolbox}

\usepackage{comment}

\usepackage{enumerate}
\usepackage{algorithm}
\usepackage{algorithmicx}
\usepackage{algpseudocode}
\usepackage{paralist}
\usepackage{threeparttable}
\usepackage{threeparttablex}
\usepackage{booktabs}
\usepackage{makecell}
\usepackage{multirow}
\usepackage{color}
\usepackage[mathscr]{eucal}
\usepackage{amsthm}
\usepackage{gensymb}
\usepackage{pifont}
\usepackage{bm}

\algnewcommand{\Inputs}[1]{
  \State \textbf{Inputs:}
  \Statex \hspace*{\algorithmicindent}\parbox[t]{.8\linewidth}{\raggedright #1}
}
\algnewcommand{\Initialize}[1]{
  \State \textbf{Initialize:}
  \Statex \hspace*{\algorithmicindent}\parbox[t]{.8\linewidth}{\raggedright #1}
}
\algnewcommand{\Outputs}[1]{
  \State \textbf{Outputs:}
  \Statex \hspace*{\algorithmicindent}\parbox[t]{.8\linewidth}{\raggedright #1}
}
\algnewcommand\algorithmicswitch{\textbf{switch}}
\algnewcommand\algorithmiccase{\textbf{case}}
\algdef{SE}[SWITCH]{Switch}{EndSwitch}[1]{\algorithmicswitch\ #1\ \algorithmicdo}{\algorithmicend\ \algorithmicswitch}%
\algdef{SE}[CASE]{Case}{EndCase}[1]{\algorithmiccase\ #1}{\algorithmicend\ \algorithmiccase}%
\algtext*{EndSwitch}%
\algtext*{EndCase}%

\newfloat{Algorithm}{htbp}{alg}
\floatname{Algorithm}{Algorithm}

\makeatletter
\def\@email#1#2{%
 \endgroup
 \patchcmd{\titleblock@produce}
  {\frontmatter@RRAPformat}
  {\frontmatter@RRAPformat{\produce@RRAP{*#1\href{mailto:#2}{#2}}}\frontmatter@RRAPformat}
  {}{}
}%
\makeatother

\UseRawInputEncoding
\begin{document}

\preprint{}

\title[Milestoning network refinement by incorporating thermodynamic and kinetic data]{Milestoning network refinement by incorporating thermodynamic and kinetic data}

\author{Xiaojun Ji}
\affiliation{Research Center for Mathematics and Interdisciplinary Sciences, Shandong University, Qingdao, Shandong 266237, P. R. China}
\affiliation{Frontiers Science Center for Nonlinear Expectations (Ministry of Education), Shandong University, Qingdao, Shandong 266237, P. R. China}

\author{Hao Wang*}
\email{wanghaosd@sdu.edu.cn}
\affiliation{
Qingdao Institute for Theoretical and Computational Sciences, School of Chemistry and Chemical Engineering, Shandong University, Qingdao, Shandong 266237, P. R. China}

\author{Wenjian Liu}
\affiliation{
Qingdao Institute for Theoretical and Computational Sciences, School of Chemistry and Chemical Engineering, Shandong University, Qingdao, Shandong 266237, P. R. China}


\begin{abstract}
Milestoning is an accurate and efficient method for rare event kinetics calculations by constructing a continuous-time kinetic network connecting the reactant and product states. However, even with adequate sampling, its accuracy can also be limited by the force fields, which makes it challenging to achieve quantitative agreement with experimental data. To address this issue, we present a refinement approach by minimizing the Kullback-Leibler divergence rate between two Milestoning networks while incorporating experimental thermodynamic (equilibrium constants) and kinetic (rate constants) data as constraints. This approach ensures that the refined kinetic network is minimally perturbed with respect to the original one, while simultaneously satisfying the experimental constraints. The refinement approach is demonstrated using the binding and unbinding dynamics of a series of six small molecule ligands for the model host system, $\beta$-cyclodextrin.
\end{abstract}

\maketitle

\onecolumngrid

\section{Introduction}
Atomistic molecular dynamics (MD) simulations is an indispensable tool for investigating important physical and chemical processes at high spatial and temporal resolution.
However, simulating long-time dynamic processes, such as protein folding, conformational change, and protein-ligand binding/unbinding, presents significant challenges for MD simulations.
Despite advancements in specialized hardware that now allow for the direct observation of mini-protein folding at milliseconds through brute-force MD simulations\cite{Shaw11}, obtaining reliable kinetic statistics remains demanding.
Accurately estimating kinetic properties typically requires observing hundreds of transition events, which requires prohibitively long simulation time.

To address this challenge, various enhanced sampling techniques for kinetics have been developed, such as Transition Interface Sampling\cite{TIS03,TIS05} (TIS), Weighted Ensemble\cite{WE96,WE10} (WE), Forward Flux Sampling\cite{FFlux06,FFlux09} (FFS), and Milestoning\cite{CM04,RevMile21}.
These methods adopt the "splitting" strategy and tackle the problem by constructing stochastic models in discretized states.
Whereas these techniques are exact for kinetics calculations, their accuracy is still limited by the underlying force fields\cite{Shaw12}.
Even with adequate sampling, the primary source of error often stems from the force fields, leading to quantitative discrepancies between simulations and experimental data. 

To integrate simulations with experimental data, various methods have been developed, such as the replica-averaged ensemble\cite{ReplicaRes05,ReplicaRes13,ReplicaRes11,ReplicaRes05-2}, Bayesian inference\cite{BayeInf14,BayeInf15}, maximum entropy\cite{MaxEnt13,MaxEnt14} (MaxEnt), and reweighting\cite{EnsReweight16,EnsReweight19}.
Among these methods, the MaxEnt approach provides a simple and general solution.
It dates back to Jaynes's seminal paper in 1957\cite{Jaynes57}.
By minimizing the relative entropy, also known as the Kullback-Liebler (KL) divergence, $S=\int d\mathbf{x}\ p(\mathbf{x})\ln\frac{p(\mathbf{x})}{p^0(\mathbf{x})}$, under the appropriate structural and thermodynamic constraints, the refined configuration distribution $p(\mathbf{x})$ is minimally perturbed from its reference form $p^0(\mathbf{x})$ while satisfying the imposed constraints.
This constrained optimization problem is usually solved using the Lagrange formalism as a posteriori analysis.
However, an equivalent replica-averaged ensemble approach can be utilized to bias the simulation on the fly.
The equivalence is exact when the harmonic potential enforcing the restraint becomes infinitely narrow and the number of replicas becomes infinitely large\cite{ResMentEqui12,ResMentEqui13,ResMentEqui13-2}. 

When the the idea of MaxEnt is extended to path ensembles, it is typically referred to as the maximum caliber\cite{MaxCal18} (MaxCal), where the relative entropy is expressed in terms of the trajectory distribution, $S=\int \mathcal{D}\mathbf{x}\  P[\mathbf{x}]\ln\frac{P[\mathbf{x}]}{P^0[\mathbf{x}]}$.
The MaxCal approach is a variational principle concerning the dynamic properties of trajectories in both equilibrium and nonequilibrium problems.
Compared to structural and thermodynamic constraints, the application of kinetic constraints has been less explored in molecular simulations, due to the more complicated form of MaxCal.
Nonetheless, recent successes have been reported in using MaxCal to reweight continuum path ensemble\cite{KinRes21,KinRes21-2} and optimize force fields\cite{KinResFF23}. 

In this work, we present an efficient way to refine the Milestoning network by imposing both thermodynamic (equilibrium constants) and kinetic (rate constants) constraints based on the MaxCal approach.
The KL divergence rate between path ensembles on two Milestoning networks is analytically evaluated and minimized as the loss function.
The refined network represents a minimal perturbation to the reference one while obeying the imposed  constraints.
This is particularly useful for correcting systematic errors stemming from force fields.

The remainder of this paper is organized as follows. 
In Sec. \ref{Methods}, we first briefly review the Milestoning formulation and then introduce the Milestoning network refinement method. 
Next, in Sec. \ref{Simulation}, computational details of the Milestoning setup for the model host-guest system and constrained optimization are summarized. 
In Sec. \ref{Results}, the refinement procedure is illustrated using the model host-guest system. 
Finally, Sec. \ref{Conclusion} contains the concluding remarks.

\section{Method}\label{Methods}
\subsection{Milestoning Method}\label{Milestoning}
Milestoning is an accurate and efficient approach for rare event kinetics calculations.
As its main kinetic output, the mean first passage time (MFPT) from a reactant $R$ and a product $P$ is calculated using a continuous-time non-Markovian stochastic model\cite{ExM15,MathExM16}.
Local short trajectory simulations are used to estimate the model parameters, transition probabilities and mean residence time.

In Milestoning, the phase space $\Gamma$ is partitioned into small compartments, e.g., by Voronoi tessellation\cite{MarkovM09}.
Interfaces between two compartments are called milestones, denoted by $\mathcal{M}=\{a,b,c,\cdots\}$.
The total number of milestones is assumed to be $N$.
The current state of a trajectory moving in $\Gamma$ is determined by the last milestone it crossed.

Consider a stationary condition, where the flux $q_a(x_a)$ across milestone $a$ is time-independent.
Here, $x_a$ denotes a phase space point on milestone $a$.
As there is no sink or source in phase space, the flux conservation is given by
\begin{equation}
q_b(x_b) = \sum_{a\in\mathcal{M}} \int dx_a q_a(x_a)K_{ab}(x_a,x_b),
\label{flux conservation}
\end{equation} 
where the summation is over all milestones from which a trajectory initiated can directly reach milestone $b$ with no other milestones being crossed in between, and $K_{ab}(x_a,x_b)$ is the average transition probability from $x_a$ to $x_b$.
The transition time from $x_a$ to $x_b$ is a random variable and has been integrated out when calculating $K_{ab}(x_a,x_b)$.

Equation \eqref{flux conservation} can be further coarse-grained by only preserving the milestone index.
To this end, the flux is split into two terms, $q_a(x_a)=q_af_a(x_a)$, where $q_a$ is the total flux across milestone $a$ and $f_a(x_a)$ is a normalized probability density function.
By integrating $x_b$ on both sides of Eq. \eqref{flux conservation}, we finally arrive at
\begin{equation}
q_b = \sum_{a\in\mathcal{M}} q_a K_{ab},
\label{flux eigequation}
\end{equation} 
where $K_{ab}=\int dx_adx_b f_a(x_a) K_{ab}(x_a,x_b)$ is the average transition probability from milestone $a$ to $b$.
For exact kinetics calculations, $f_a(x_a)$ should be the first hitting point distribution\cite{ExM15} (FHPD), which corresponds to the phase space points that a long trajectory first hits milestone $a$ after crossing a different milestone. 
It is frequently approximated as the Boltzmann distribution for simplicity\cite{CM04,Wang21}, $f_a(x_a)\sim\exp(-\beta H(x_a))$.
But more accurate approximations are available at a higher cost\cite{LPTM,DiM10,BuffM,ExM15}.

To efficiently estimate $K_{ab}$, we sample $n_a$ initial configurations according to $f_a(x_a)$ on milestone $a$ and run free trajectories until they hit a different milestone,
\begin{equation}
K_{ab}=\frac{n_{ab}}{n_a},
\label{K estimation}
\end{equation}
where $n_{ab}$ is the number of $a\rightarrow b$ transitions.
In addition, the mean residence time on each milestone can be estimated as
\begin{equation}
\bar{t}_a = \frac{\sum_{l=1}^{n_a} t_a(l)}{n_a},
\label{t estimation}
\end{equation}
where $t_a(l)$ is the lifetime of the $l$-th trajectory initiated from milestone $a$.

Finally, the MFPT is calculated by
\begin{equation}
\bm{\tau} = (\mathbf{I}-\mathbf{K}^{(A)})^{-1}\mathbf{\bar{t}},
\label{mfpt Kt}
\end{equation} 
where $\bm{\tau}$ is a column vector with its element being the MFPT from each state to the product state $P$, $\mathbf{I}$ is an identity matrix, and elements of $\mathbf{K}^{(A)}$ and $\mathbf{\bar{t}}$ are calculated via Eqs. \eqref{K estimation} and \eqref{t estimation}, respectively.
Note that the absorbing boundary condition is set at the product $P$ in Eq. \eqref{mfpt Kt}, i.e., $K^{(A)}_{Pa}=0$, $\forall a\in\mathcal{M}$ and $\bar{t}_P=0$. 
In particular, the MFPT of the $R\rightarrow P$ transition is given by
\begin{equation}
\tau_{R\rightarrow P}=\mathbf{e}_R^T\bm{\tau},
\end{equation}
where $\mathbf{e}_R$ a unit vector with the $R$-th element being one and all other elements being zero.

In addition, the stationary probability of the last crossed milestone being $a$ is given by
\begin{equation}
\pi_a = q_a \bar{t}_a,
\label{pi def}
\end{equation}
where the stationary flux is solved from Eq. \eqref{flux eigequation}.

\subsubsection*{A Markovian Equivalent}
The Milestoning formalism is an exact approach for MFPT calculations.
The non-Markovian effect is implicitly contained in the FHPD, and the transition times between two milestones are in general not exponentially distributed.
However, an equivalent continuous-time Markov chain (CTMC) that shares the same stationary probabilities, mean residence time, and MFPT as the original Milestoning formalism can be constructed\cite{Elber07,CTMC19}. 
The CTMC is described by a single transition rate matrix $\mathbf{Q}$,
\begin{align}
Q_{ab}&=K_{ab}/\bar{t}_a\ \mathrm{for}\ a\neq b,\nonumber\\
Q_{aa}&=-\sum_{b\neq a}Q_{ab}=-1/\bar{t}_a.
\label{Q def}
\end{align} 

In the CTMC formalism, the stationary probability is given by
\begin{equation}
\bm{\pi}^T\mathbf{Q}=\mathbf{0}^T,
\label{pi equ}
\end{equation}
where $\mathbf{0}^T=(0,\cdots,0)$ is a zero vector of $N$ elements.
The MFPT of the $R\rightarrow P$ transition is given by
\begin{equation}
\tau_{R\rightarrow P}=-\mathbf{e}_R^T\mathbf{Q}'^{-1}\mathbf{1},
\label{mfpt Q}
\end{equation}
where $\mathbf{Q}'$ is a $(N-1)\times(N-1)$ matrix with the $P$-th row and column being deleted from $\mathbf{Q}$, and $\mathbf{1}$ is a column vector of $(N-1)$ ones.
Eq. \eqref{pi equ} is essentially a combination of Eqs. \eqref{flux eigequation} and \eqref{pi def}, and Eq. \eqref{mfpt Q} can be derived from Eq. \eqref{mfpt Kt}.

This equivalent CTMC serves as a convenient basis for the Milestoning network refinement discussed in the next section.  

\subsection{Milestoning Network Refinement}\label{Refinement}
\subsubsection{KL Divergence Rate between Two Milestoning Networks}\label{RelEnt}
Consider the time evolution of a trajectory $y(t)$ on an ergodic CTMC, denoted by $\omega=(t_0,y_0,t_1,y_1,\cdots,t_{n},y_{n})$, where $t_i$ is the first hitting time on $y_i\in\mathcal{M}$ after crossing a different milestone.
The probability density function associated with such a trajectory is given by
\begin{equation}
\mathcal{P}_n(\omega) = \rho(y_0)\Pi_{i=0}^{n-1}Q_{y_iy_{i+1}}\exp(Q_{y_iy_i}(t_{i+1}-t_i)),
\end{equation}
where $\rho(y_0)$ is the initial distribution.

For path ensemble evolving on two different CTMCs, $\mathbf{Q}$ and $\mathbf{Q}^0$, the KL divergence between these two path distribution is given by
\begin{equation}
D(\mathcal{P}_n||\mathcal{P}_n^0) = \sum_\omega \mathcal{P}_n(\omega)\ln\frac{\mathcal{P}_n(\omega)}{\mathcal{P}_n^0(\omega)},
\end{equation} 
which measures the distance between these two path distributions.
After some algebra (see Supporting Information), the analytic expression for $D(\mathcal{P}_n||\mathcal{P}_n^0)$ is obtained.
It is found that the KL divergence $D(\mathcal{P}_n||\mathcal{P}_n^0)$ is path length dependent.
To get rid of this dependency, the KL divergence rate in the limit of infinitely long trajectories is defined as follows,
\begin{align}
D(\mathbf{Q}||\mathbf{Q}^0) &= \lim_{n\rightarrow\infty}\frac{1}{n}D(\mathcal{P}_n||\mathcal{P}_n^0)\nonumber\\
&= \sum_{a\in\mathcal{M}}\pi_a(\sum_{b\neq a}Q_{ab}\ln\frac{Q_{ab}}{Q_{ab}^0}+Q_{ab}^0-Q_{ab}).
\label{D def}
\end{align}
$D(\mathbf{Q}||\mathbf{Q}^0)$ has the property that
\begin{equation}
D(\mathbf{Q}||\mathbf{Q}^0)\geq 0,
\end{equation}
where the equality holds if and only if $\mathbf{Q}=\mathbf{Q}^0$.
Therefore, $D(\mathbf{Q}||\mathbf{Q}^0)$ is used as the loss function for the constrained optimization in the next section.
In other words, the refined $\mathbf{Q}$ is kept as close to the original $\mathbf{Q}^0$ as possible.

\subsubsection{Constrained Optimization}
Both thermodynamic (equilibrium constants) and kinetic (rate constants) constraints are considered in this work. 
For the protein-ligand binding and unbinding processes, the association and dissociation rate constants, $k_{on}$ and $k_{off}$, are related to the MFPT via
\begin{align}
k_{on} &= \frac{1}{\tau_{on}[L]},\\
k_{off} &= \frac{1}{\tau_{off}},
\end{align}
where $[L]$ is the ligand concentration.
The corresponding equilibrium constant is called the binding constant, 
\begin{equation}
K_a=k_{on}/k_{off},
\label{Ka def}
\end{equation}
and can be used to evaluate the binding free energy, 
\begin{equation}
\Delta G = -RT\ln(C_0K_a),
\end{equation}
where $C_0$ is the standard concentration 1 molar.

Therefore, the constrained optimization problem is summarized as follows,
\begin{align}
&\min_{\mathbf{Q}}\  D(\mathbf{Q}||\mathbf{Q}^0),\\
\mathrm{subject\ to:} &\ Q_{ab} \geq 0, \mathrm{for}\ a\neq b\in \mathcal M,\label{bound cons}\\
&Q_{aa}=-\sum_{b\neq a}Q_{ab}, \mathrm{for}\ a\in\mathcal{M},\label{diag cons}\\
&\ln(k_{on}^{exp}-\sigma_{on}^{exp})\leq c_1(\mathbf{Q})=\ln k_{on}\leq\ln(k_{on}^{exp}+\sigma_{on}^{exp}),\label{c1 func}\\
&\ln(k_{off}^{exp}-\sigma_{off}^{exp})\leq c_2(\mathbf{Q})=\ln k_{off}\leq\ln(k_{off}^{exp}+\sigma_{off}^{exp}),\label{c2 func}\\
&\ln(K_a^{exp}-\sigma_a^{exp})\leq c_3(\mathbf{Q})=\ln K_a\leq\ln(K_a^{exp}+\sigma_a^{exp}),\label{c3 func}
\end{align}
where the superscript "exp" indicates the experimental data and $\sigma$ denotes the statistical uncertainty.
Equations. \eqref{bound cons} and \eqref{diag cons} are the natural constraints for transition rate matrices.
The number of thermodynamic and kinetic constraints involved (Eqs. \eqref{c1 func}-\eqref{c3 func}) depends on the availability of experimental data. 

\subsubsection{Explicit Constraint on Mean Residence Time}
The loss function $D(\mathbf{Q}||\mathbf{Q}^0)$ aims to minimize the discrepancy between the refined and original transition rate matrix, $\mathbf{Q}$ and $\mathbf{Q}^0$.
As can be seen from the definition of $D(\mathbf{Q}||\mathbf{Q}^0)$ in Eq. \eqref{D def}, only non-diagonal elements of $\mathbf{Q}$ and $\mathbf{Q}^0$ are explicitly involved.
The discrepancy between diagonal elements is only indirectly constrained by Eq. \eqref{diag cons}.

The diagonal elements of $\mathbf{Q}$ is related to the mean residence time via Eq. \eqref{Q def}.
In practice, we find it helpful to stabilize the optimization by explicitly constraining the diagonal elements.
To this end, we assume the lifetime of each sampled trajectory initiated from milestone $a$ is independent and identically distributed.
By the central limit theorem, the distribution of $\bar{t}_a$ obeys a normal distribution, $f(\bar{t}_a)\sim\mathcal{N}(\mu_{\bar{t}_a},\sigma_{\bar{t}_a}^2)$.
Here, the average of $\bar{t}_a$, $\mu_{\bar{t}_a}$, is the same as $\bar{t}_a$ and is estimated from the sample via Eq. \eqref{t estimation}.
The standard deviation of $\bar{t}_a$ is estimated by
\begin{align}
\sigma_{\bar{t}_a} &= \sigma_{t_a}/\sqrt{n_a},\\
\sigma_{t_a} &= \sqrt{\frac{\sum_{l=1}^{n_a}(t_a(l)-\bar{t}_a)^2}{n_a-1}}, 
\end{align}
where $\sigma_{t_a}$ is the standard deviation of a trajectory's lifetime $t_a(l)$.

Finally, the explicit constraint for the diagonal elements is expressed as
\begin{equation}
\frac{1}{\mu_{\bar{t}_a}+\sigma_{\bar{t}_a}}\leq -Q_{aa}\leq \frac{1}{\mu_{\bar{t}_a}-\sigma_{\bar{t}_a}}, \mathrm{for}\ a\in\mathcal{M}.
\label{linear cons}
\end{equation} 

\section{Computational Details}\label{Simulation}
\subsection{Simulation Setup}
\begin{figure}[h]
\centering
\begin{tabular}{cccccc}
\makecell{
\includegraphics[height=5cm]{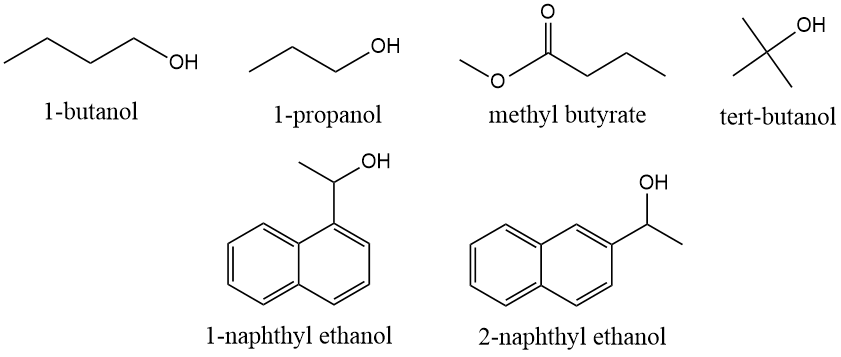} \\
(a)}\\
\makecell{
\includegraphics[height=8cm]{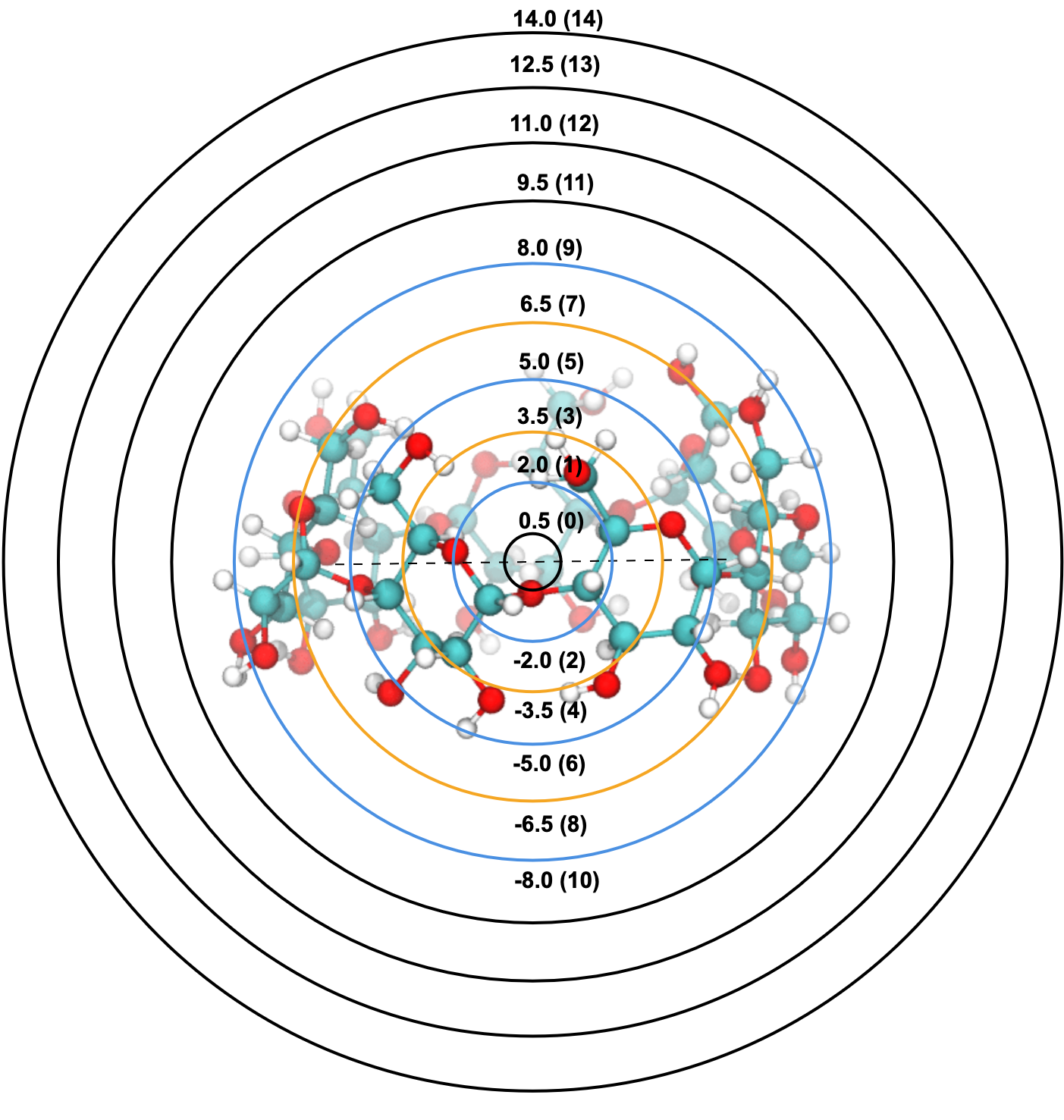} \\
(b)}\\
\end{tabular}
\caption{(a) Structures for six small molecule ligands. (b) Spherical milestones are defined between the center of mass of $\beta$-CD and that of a ligand. Blue and yellow milestones are further divided into two half-spheres. The numbers in parenthesis denote milestone indices. The innermost (milestone 0) and outermost (milestone 14) milestones are defined as the bound and unbound states, respectively.}\label{fig_milestoning}
\end{figure}

Four weak binders (1-butanol, 1-propanol, methyl butyrate and tert-butanol) and two relatively strong binders (1-naphthyl ethanol and 2-naphthyl ethanol) are selected as the guest molecules for $\beta$-cyclodextrin ($\beta$-CD) (Fig. \ref{fig_milestoning} (a)).
CGenFF force fields are used for $\beta$-CD and all six ligands\cite{CGenFF}.
The program NAMD 2.14 is used for all simulations\cite{NAMD}.
Each host-guest complex is solvated in TIP3P water molecules.
Periodic boundary conditions are applied in all three directions.
The system is minimized with a conjugate gradient algorithm for 10000 steps and then is equilibrated at 298 K for 250 ps.
A longer equilibration is performed in NPT ensemble using a Nose-Hoover Langevin piston pressure control\cite{NPT94,NPT95} at 298 K and 1 atm for 10 ns before Milestoning simulations. 

In all simulations, water molecules are kept rigid using the SETTLE algorithm\cite{SETTLE}, and all other bond lengths with hydrogen atoms are kept fixed using the SHAKE algorithm\cite{SHAKE}. 
The integration time step is 2 fs. 
A real space cutoff distance of 10 \r{A} is used for both electrostatic and van der Waals interactions.
Particle mesh Ewald is utilized for long-range electrostatic calculations\cite{PME}.

The configuration space is partitioned based on the distance between the center of mass (COM) of $\beta$-CD and that of a guest molecule (see Fig. \ref{fig_milestoning} (b)).
Spherical milestones are defined from 0.5 \r{A} to 14 \r{A} with an interval of 1.5 \r{A}.
The spherical milestones at 0.5 \r{A} and 14 \r{A} are defined as the bound and unbound states, respectively.
Considering the asymmetry of $\beta$-CD, spherical milestones within 8 \r{A} except the bound state are further divided into two half-spheres.
When the COM distance is larger than 8 \r{A}, guest molecules can freely sample both faces.
Therefore, a single surface is enough for initial configuration sampling.
As a result, there are a total number of 15 milestones.

To obtain initial configurations, a restrained sampling in the NVT ensemble at 298 K is performed on each milestone with a force constant of 90 kcal/mol/\r{A}${^2}$.
The restrained simulation runs 60 ns for four weak binders and 110 ns for two strong binders.
The first 10 ns simulation is discarded for equilibration, and configurations are saved every 0.1 ns.
As a result, 500 configurations per milestone are prepared for four weak binders and 1000 configurations per milestone are prepared for two strong binders.
Free trajectories evolved from sampled configurations are used for Milestoning calculations.
Bootstrapping analysis is performed for standard deviation estimation.

\subsection{Constrained Optimization}
The constrained optimization is performed using the trust-region constrained algorithm implemented in {}SciPy\cite{ConsOpt99}, in which constraints are divided into three classes: bound constraints (Eq. \eqref{bound cons}), linear constraints (Eq. \eqref{linear cons}) and nonlinear constraints (Eqs. \eqref{c1 func}-\eqref{c3 func}).
Even though experimental data for $k_{on}$, $k_{off}$ and $K_a$ are all available, only two out of them are independent.
Therefore, only two nonlinear constraints (Eqs. \eqref{c2 func} and \eqref{c3 func}) are finally used in practice. 
The constraint Eq. \eqref{diag cons} is directly enforced when constructing $\mathbf{Q}$.

Only nonzero non-diagonal elements of $\mathbf{Q}^0$ are optimized.
The gradient and hessian information of the loss function ($D(\mathbf{Q}||\mathbf{Q}^0)$) and nonlinear constraints required for optimization is provided in the Supporting Information. 

\section{Results and Discussion}\label{Results}
The model host-guest system, $\beta$-CD with a series of six ligands of various binding affinities, is used for illustration purpose.
The relatively small size of this system allows for extensive sampling of initial configurations and transition trajectories.
The mean residence time at each milestone separated by 1.5 \r{A} ranges from several to hundreds of picoseconds, which is much longer than the typical velocity decorrelation time at subpicoseconds. 
Consequently, the sampling errors associated with the initial distribution and transition events are substantially reduced.
Therefore, the errors in kinetic ($k_{on}$ and $k_{off}$) and thermodynamic ($\Delta G$) estimations mainly reflect the quality of the force field used. 
This model system has been extensively studied with various force fields, including GAFF and q4MD\cite{HGModel18,HGModel18-2}.
The quantitative discrepancies between the simulations and experimental data observed in our study below align with findings from these previous studies, which reveals the challenges in parameterizing force fields for small molecules.

\begin{figure}[h]
\centering
\begin{tabular}{cccccc}
\makecell{
\includegraphics[height=6.5cm]{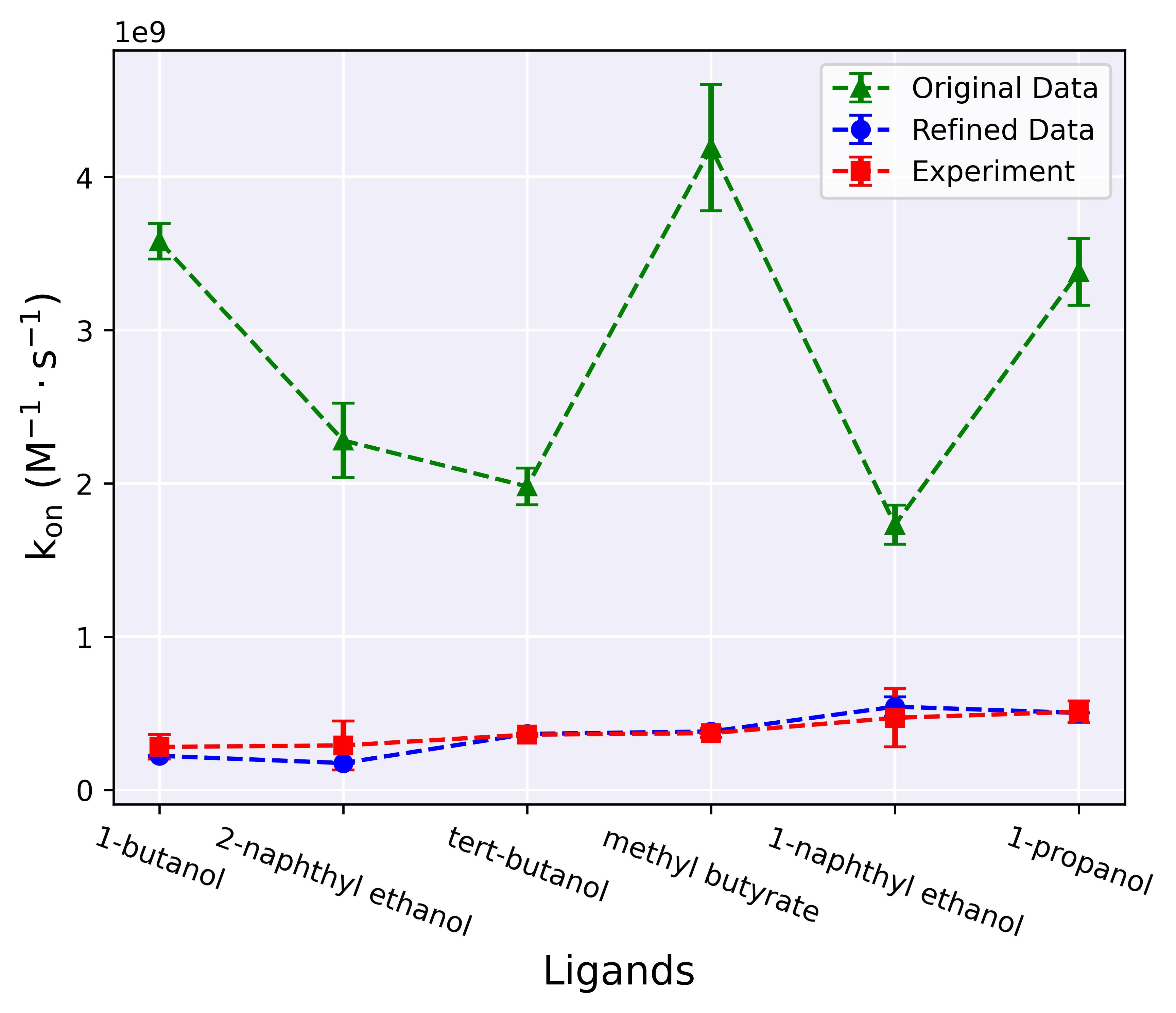} \\
(a)}\\
\makecell{
\includegraphics[height=6.5cm]{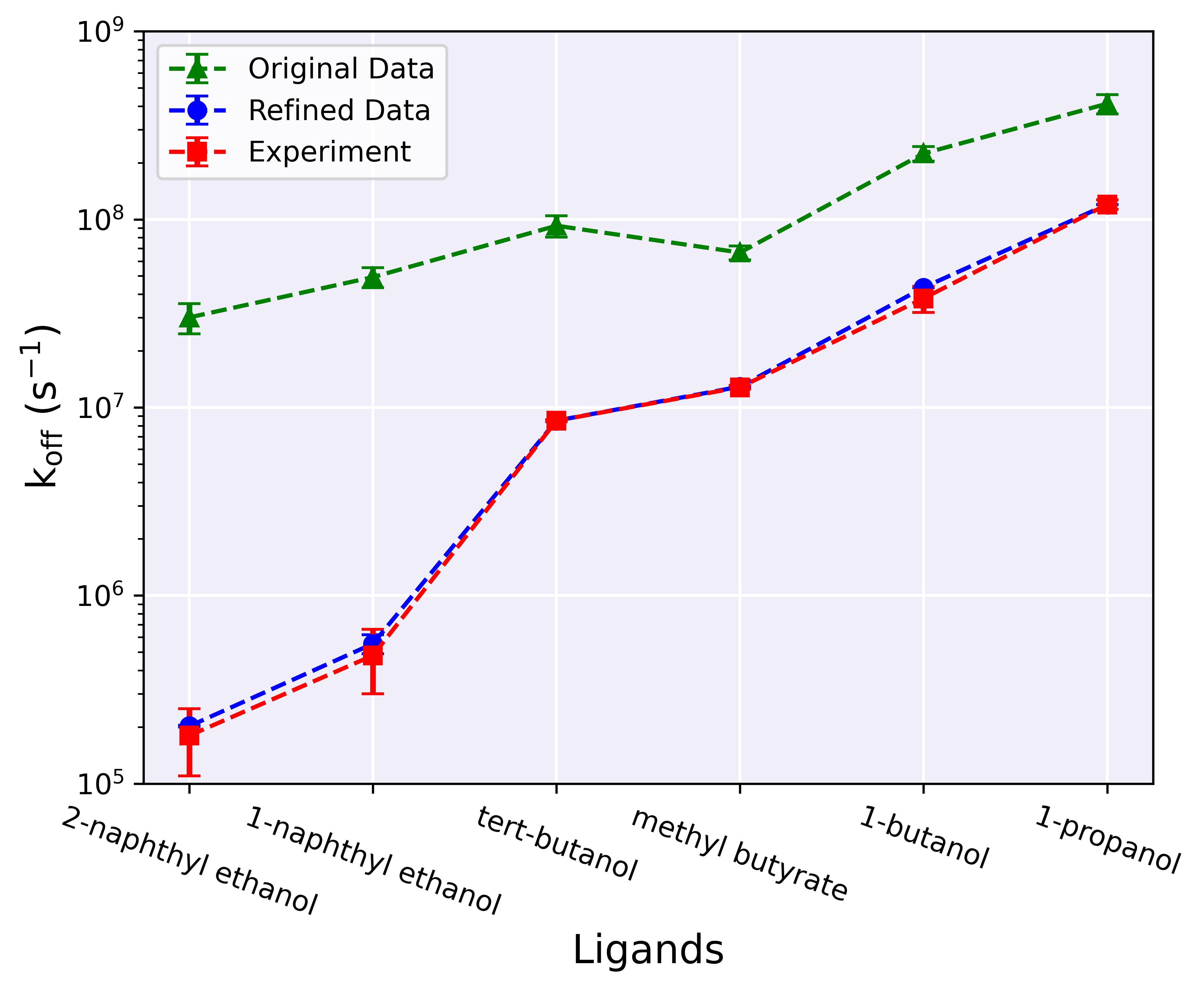} \\
(b)}\\
\makecell{
\includegraphics[height=6.5cm]{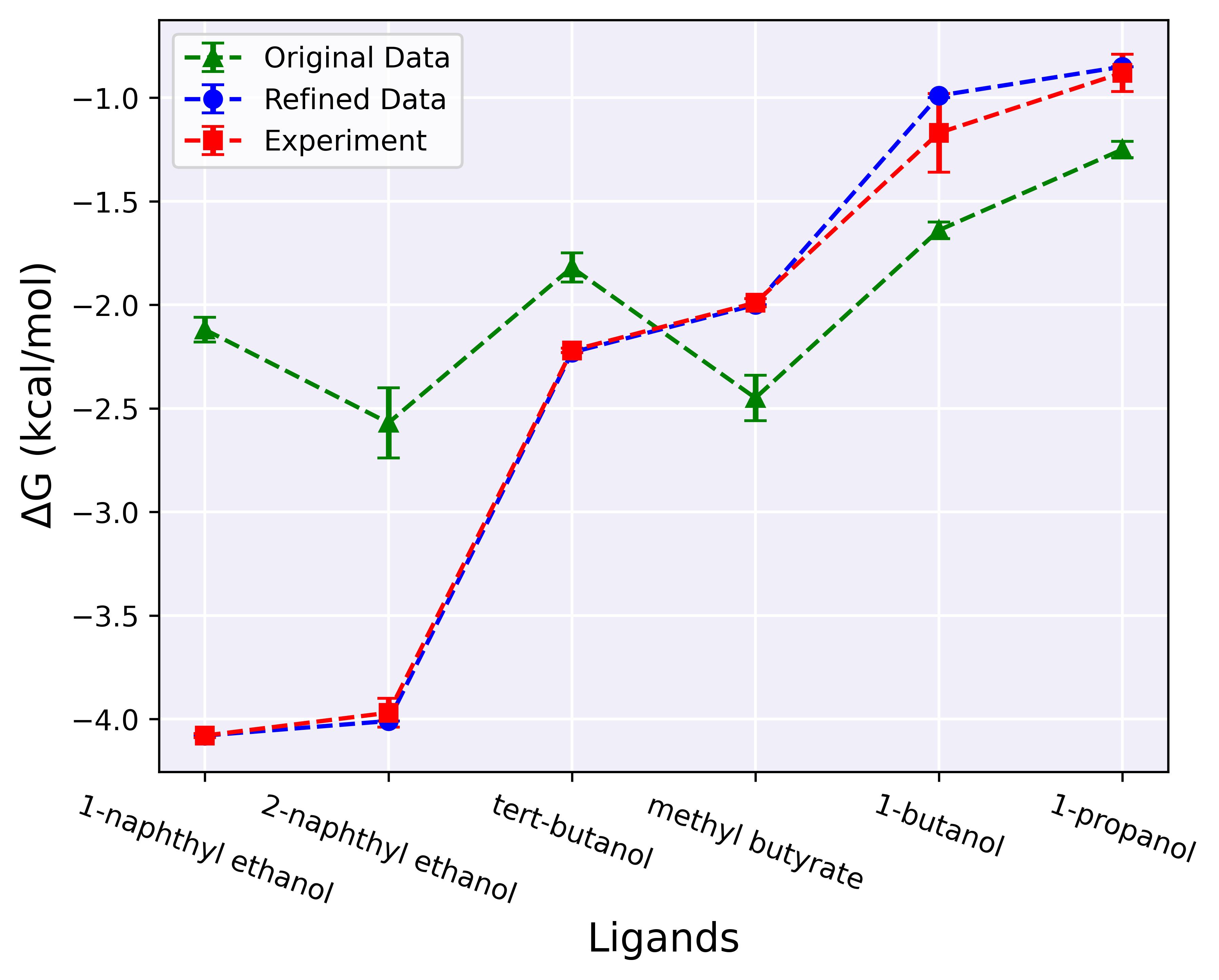} \\
(c)}
\end{tabular}
\caption{(a) Binding rate constants $k_{on}$, (b) unbinding rate constants $k_{off}$, and (c) binding free energies $\Delta G$ for six ligands.}\label{fig_kG}
\end{figure}

As shown in Fig. \ref{fig_kG} (a), the six ligands show similar experimental binding rates $k_{on}$.
Although the absolute errors in the original simulations are all around or less than one order of magnitude, they fail to quantitatively rank the ligands by increasing $k_{on}$.
The Spearman correlation coefficient between the original simulation and experimental data is as low as $-0.18$, which verifies this observation.
While $k_{on}$ is not explicitly constrained during optimization, it is related to $k_{off}$ and $K_a$ via Eq. \eqref{Ka def}, leading to the refined $k_{on}$ data that agree well with the experiments.

In contrast to $k_{on}$, the experimental $k_{off}$ data span multiple orders of magnitude, reflecting diverse residence time in the bound state (Fig. \ref{fig_kG} (b)).
The absolute errors of the simulated $k_{off}$ are more significant.
However, the simulation can correctly rank most ligands by their residence time, which is also indicated by a Spearman correlation coefficient of $0.94$.
By including the constraint of $k_{off}$ (Eq. \eqref{c2 func}) into the optimization process, the absolute errors of $k_{off}$ are significantly reduced.

The binding free energy $\Delta G$ is mainly affected by $k_{off}$ rather than $k_{on}$, since $k_{on}$ values are similar across all six ligands.
The absolute errors in $\Delta G$ for the two strong binders are notably underestimated, making it difficult to distinguish them from the weak binders (Fig. \ref{fig_kG} (c)).
The corresponding Spearman correlation coefficient is about $0.68$.
At no surprise, after optimization with the $K_a$ constraint (Eq. \eqref{c3 func}), the absolute $\Delta G$ agrees well with the experimental data.

\begin{figure}[h]
\centering
\begin{tabular}{cccccc}
\makecell{
\includegraphics[height=6cm]{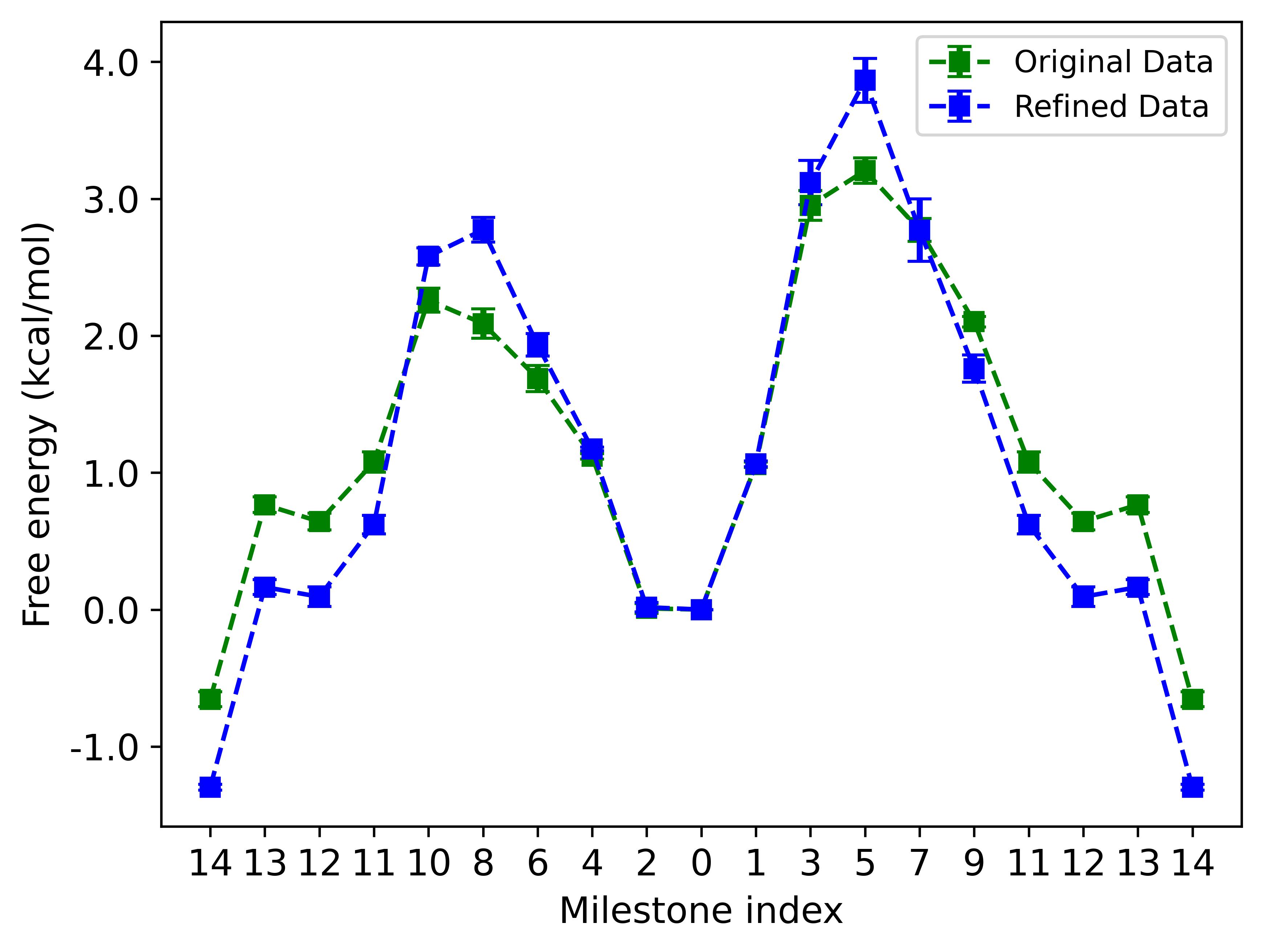}\\
(a)} &
\makecell{
\includegraphics[height=6cm]{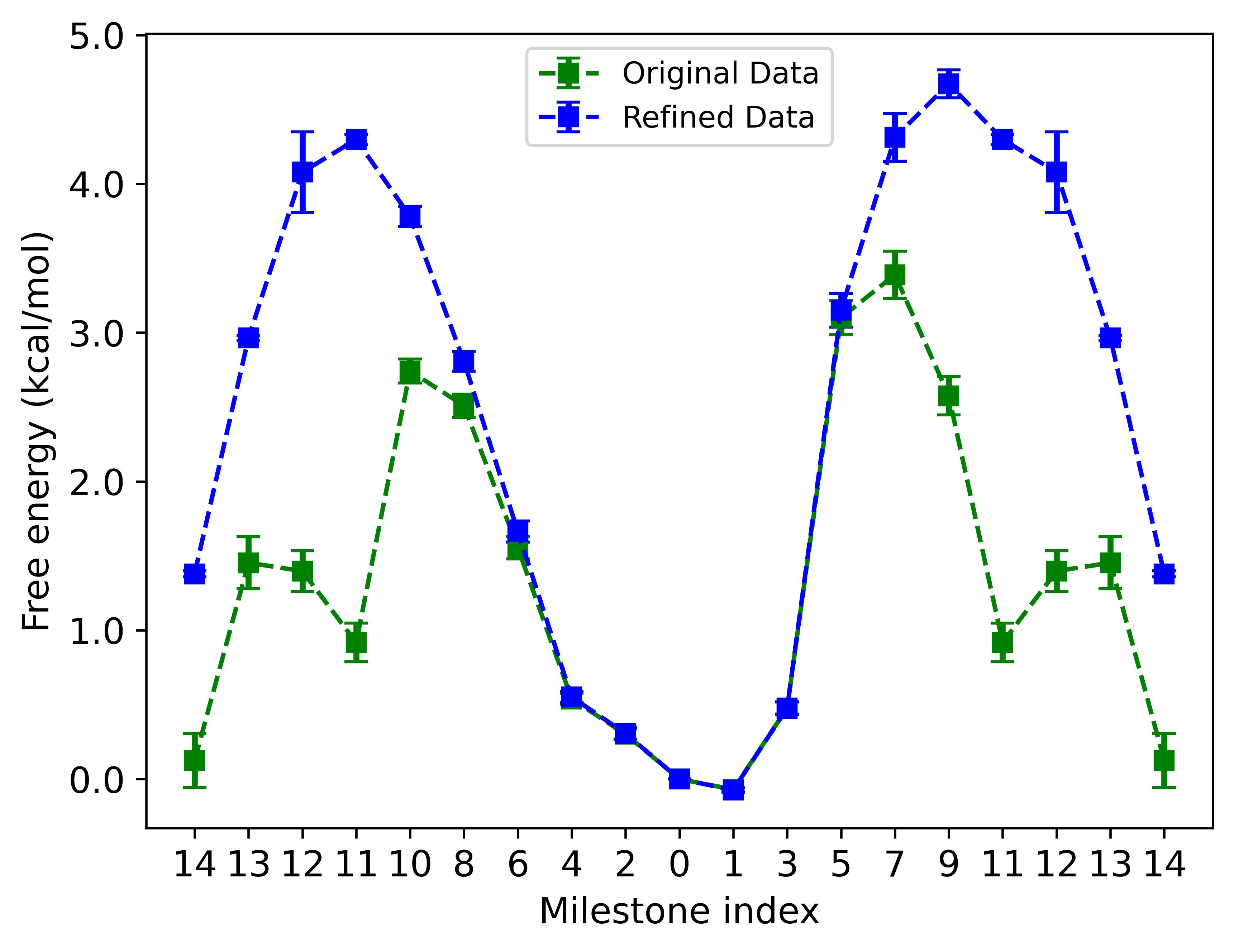}\\
(b)}\\
\makecell{
\includegraphics[height=6cm]{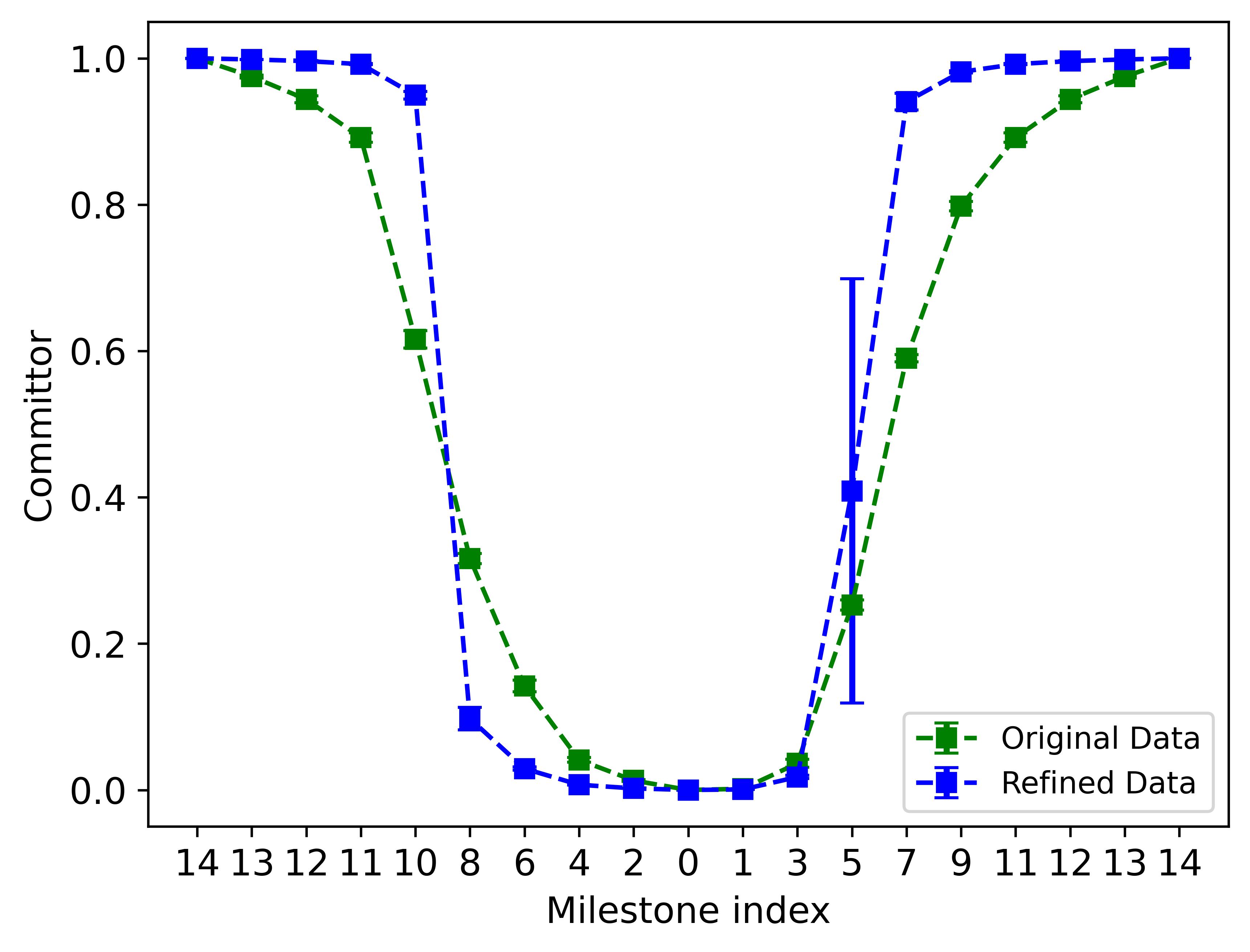} \\
(c)
} &
\makecell{
\includegraphics[height=6cm]{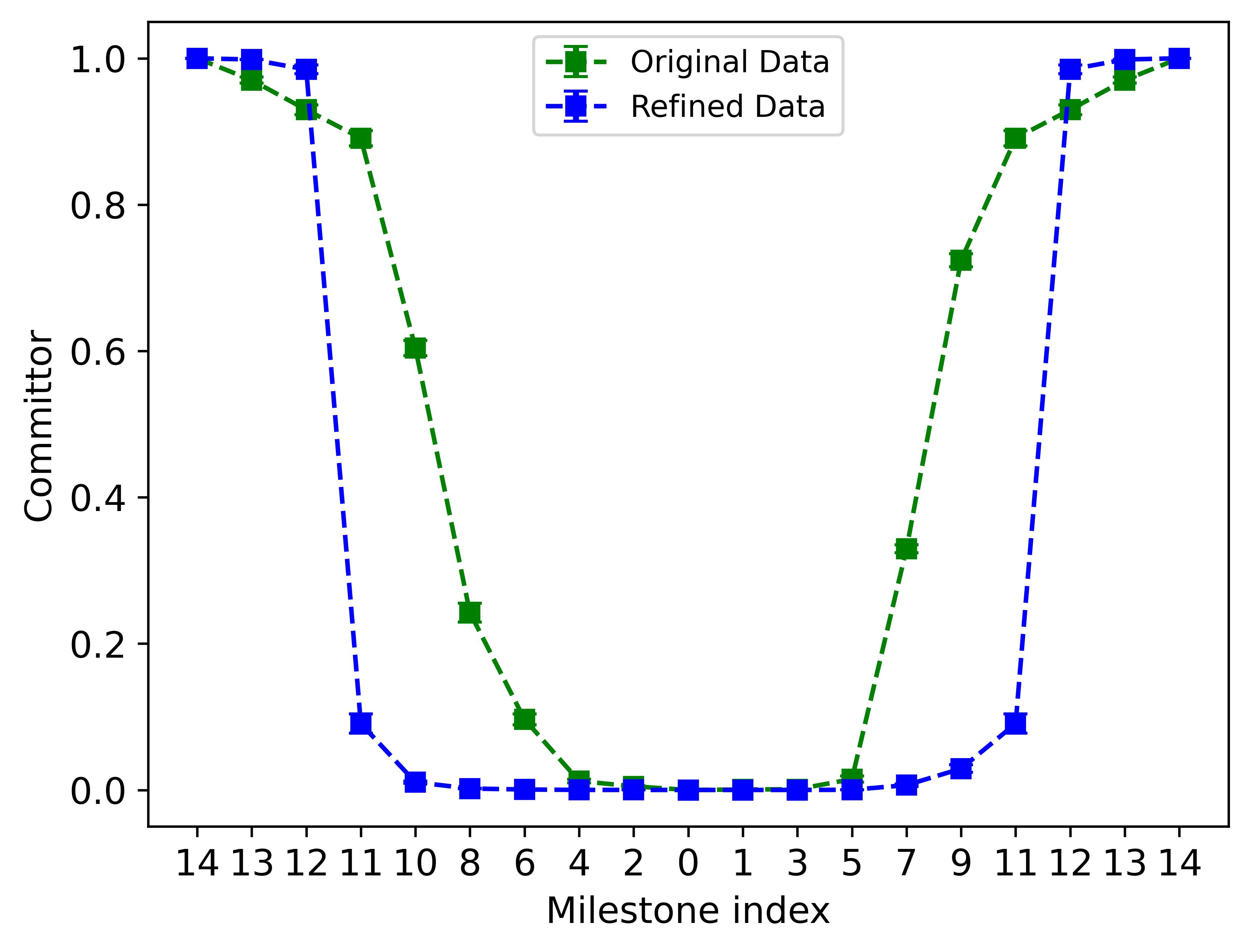} \\
(d)
}
\end{tabular}
\caption{Free energy profiles for (a) 1-butanol and (b) 2-naphthyl ethanol. Committor functions for (c) 1-butanol and (d) 2-naphthyl ethanol.}\label{fig_FC}
\end{figure}

In addition to the binding free energy (i.e., the free energy difference at two ends), the free energy profile provides a more detailed picture (Fig. \ref{fig_FC} and Supporting Information).
The activation energy along the binding and unbinding directions significantly affects $k_{on}$ and $k_{off}$, respectively.
Although the simulated $k_{on}$ and $k_{off}$ are consistently faster than experimental data for all six ligands, the most pronounced errors occur in the estimation of $k_{off}$ for the two strong binders.
Therefore, it is not surprising to find that the activation energy barrier along the unbinding direction for these two strong binders changes the most after refinement.
This increased activation energy barrier along the unbinding direction also delays the occurrence of the transition state (committor $=0.5$) for two strong binders.
Four the four weak binders, the errors in $k_{on}$ and $k_{off}$ estimations are relatively small.
Therefore, the resulting changes in the free energy profile and committor functions are modest.
However, all these changes lead to a better alignment with experimental rate constants.

\begin{figure}[h]
\centering
\begin{tabular}{cccccc}
\makecell{
\includegraphics[height=7cm]{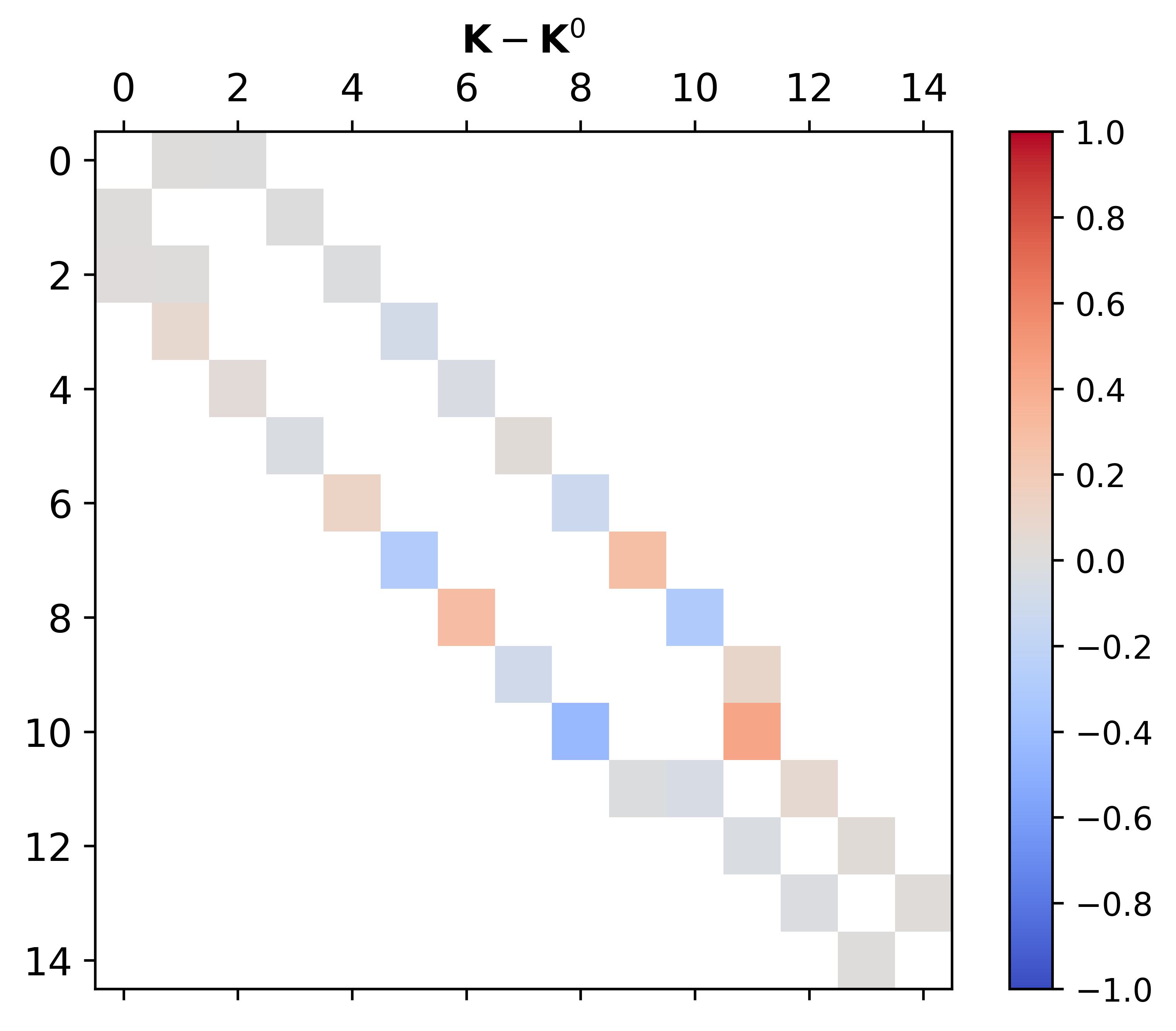}\\
(a)} &
\makecell{
\includegraphics[height=7cm]{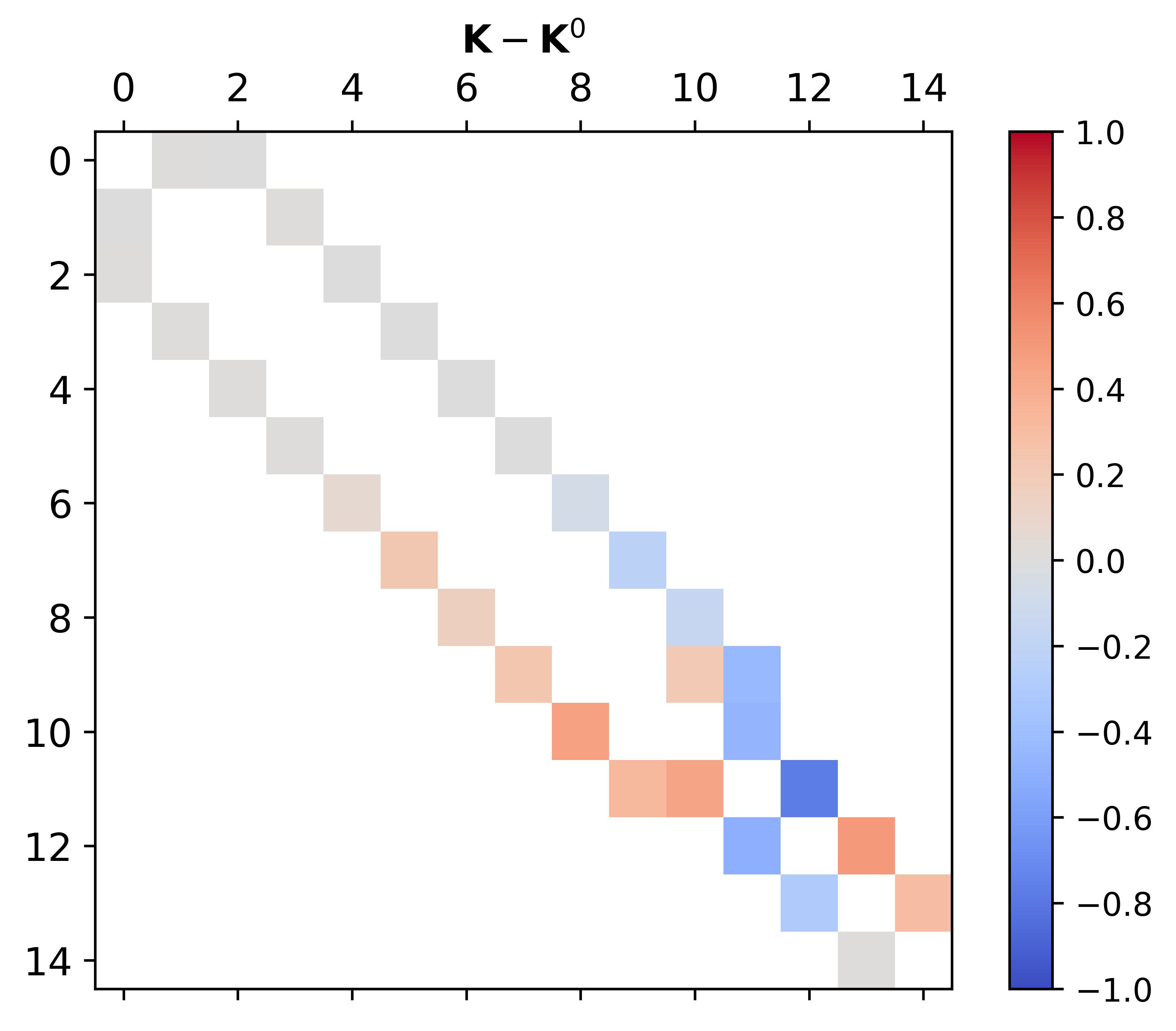}\\
(b)}\\
\makecell{
\includegraphics[height=6cm]{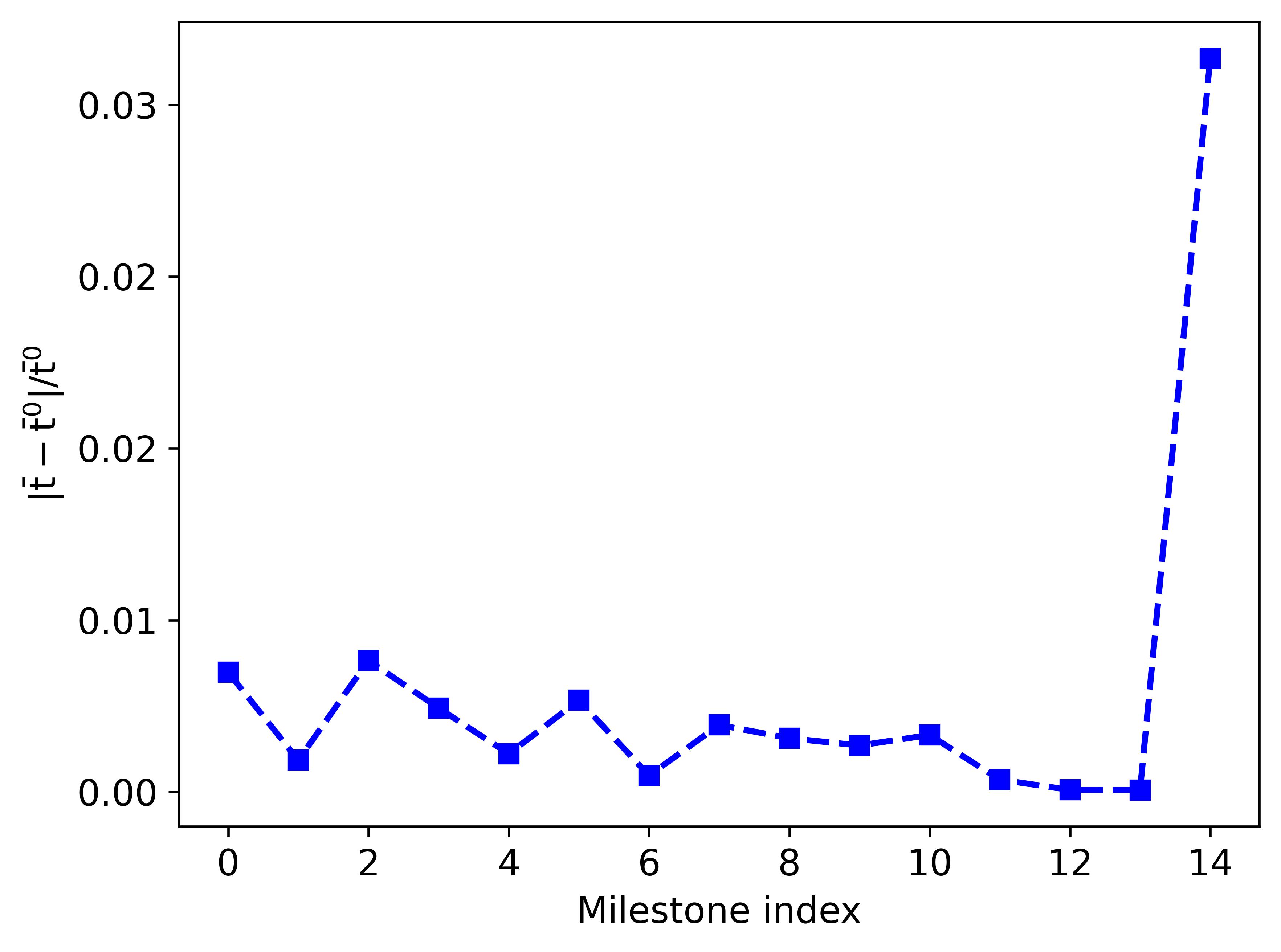} \\
(c)
} &
\makecell{
\includegraphics[height=6cm]{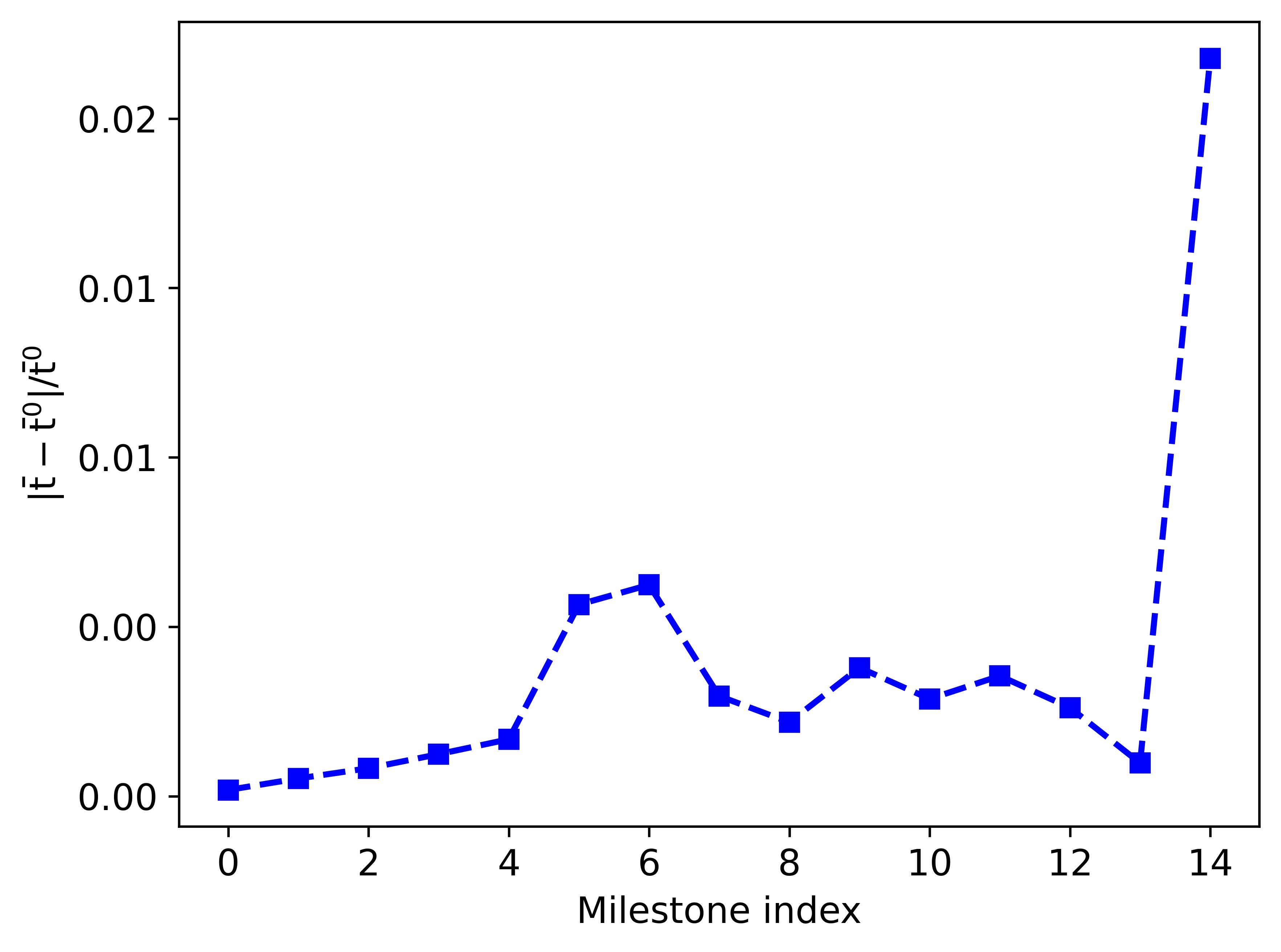} \\
(d)
}
\end{tabular}
\caption{Transition probability matrix differences for (a) 1-butanol and (b) 2-naphthyl ethanol. Relative errors of the mean residence time for (c) 1-butanol and (d) 2-naphthyl ethanol.}\label{fig_err}
\end{figure}

To assess how the refinement procedure affects the transition probabilities $\mathbf{K}$ and the mean residence time $\mathbf{\bar{t}}$, we transform the refined transition rate matrix $\mathbf{Q}$ back into $\mathbf{K}$ and $\mathbf{\bar{t}}$ by inverting Eq. \eqref{Q def} and compare them with the original simulation data (see Fig. \ref{fig_err} and Supporting Information).
Given that the standard deviation of the mean residence time, $\sigma_{\bar{t}_a}$, is typically one order of magnitude smaller than the average itself $\mu_{\bar{t}_a}$, the explicit constraint concerning the mean residence time (Eq. \eqref{linear cons}) results in a small relative deviation in $\mathbf{\bar{t}}$. 
Therefore, the refinement is mainly reflected in $\mathbf{K}$.
Note that both the simulated $k_{on}$ and $k_{off}$ are faster than the experimental data.
To align with the experimental $k_{on}$, the transition probabilities along the binding direction need to decrease.
In other words, this implies that the transition probabilities along the unbinding direction should increase.
However, to conform with experimental $k_{off}$, the transition probabilities along the unbinding direction need to decrease.
Consequently, the refined $\mathbf{K}$ results from the counteraction of these two opposing factors.

\section{Conclusion}\label{Conclusion}
We present a method to refine a given Milestoning network by incorporating thermodynamic and kinetic experimental data.
Our method is based on the MaxCal approach, ensuring that the refined network represents a minimal perturbation to the original one.
The KL divergence rate between two Milestoning networks is analytically evaluated and  used as the loss function.
We demonstrate the application of this approach on a model host system with a series of small molecule ligands, which exhibit qualitative errors in $k_{on}$ and significant quantitative errors in $k_{off}$.
The refined network shows alterations in the free energy profile and transition state, aligning more closely with experimental data.

The presented method is useful when the force field accuracy is limited or when new experimental data become available after the simulation is done.
This scenario is especially relevant in drug discovery, where high accuracy force fields for small molecule ligands are often lacking due to their diversities.
Furthermore, integrating simulations and experimental data helps identifying the correct transition state structure, which is important for the lead optimization\cite{Shaw13}.

\clearpage
\newpage

\begin{acknowledgments}
The work was partially supported by Natural Science Foundation of China (No. 22403056 and No. 12401179), Qilu Young Scholars Program of Shandong University, and Natural Science Foundation of Shandong Province (No. ZR2022QA012).
\end{acknowledgments}

\section*{Data Availability Statement}
The data that support the findings of this study are available within the article and its supplementary material.

\section*{Conflicts of interest}
There are no conflicts to declare.

\section*{Supporting Information}
Derivation of Eq. \eqref{D def}; Gradient and hessian of the loss function and constraints; Free energy profiles, committor functions, transition probability matrix differences, and relative errors of the mean residence time for 1-propanol, methyl butyrate, tert-butanol, and 1-naphthyl ethanol, respectively.

\clearpage
\newpage



\bibliographystyle{achemso1}
\bibliography{RefineM}

\clearpage
\newpage

TOC
\begin{figure}[h]
\centering
\begin{tabular}{cc}
\includegraphics[height=5cm]{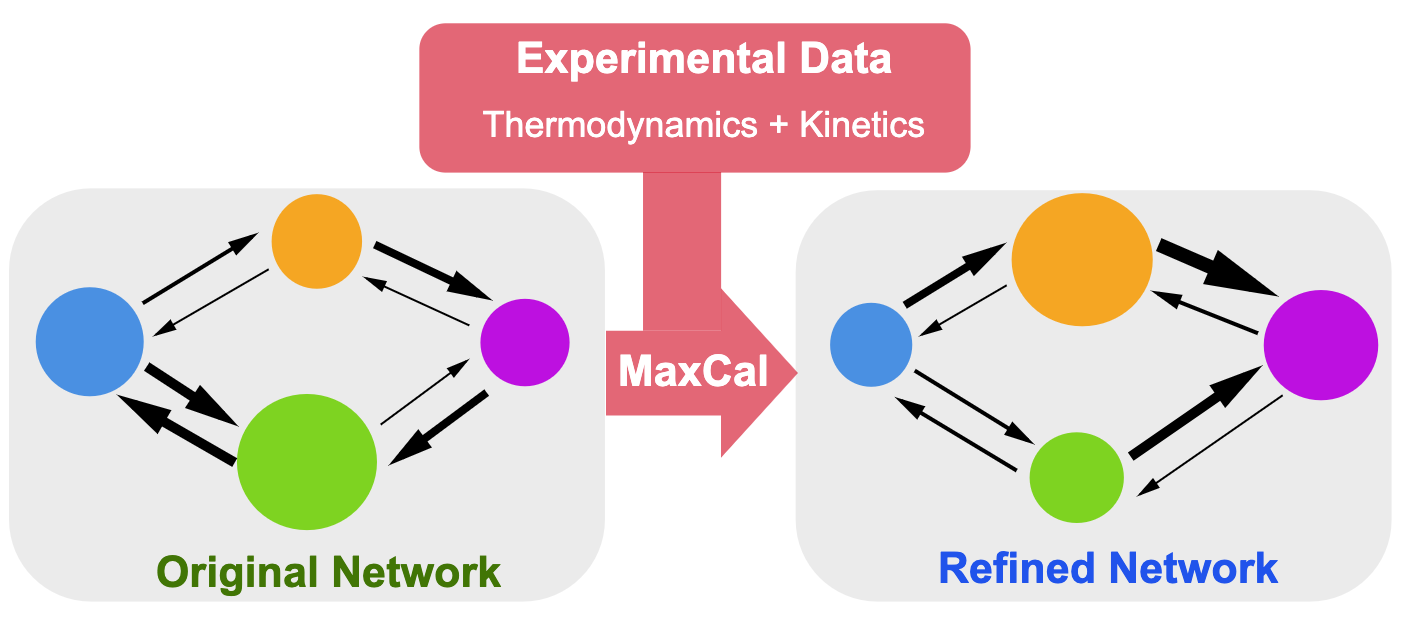}\\
\end{tabular}
\end{figure}

\end{document}